\documentclass[%
 reprint,
 amsmath,amssymb,
 aps]{revtex4-1}

\usepackage{amssymb}
\usepackage[dvipdfmx]{graphicx}
\usepackage{dcolumn}
\usepackage{bm}
\usepackage{braket}
\usepackage{color}
\usepackage{here}
\usepackage{mathtools} 
\usepackage{comment}

\begin{document}

\def\bra#1{\langle{#1}|}
\def\ket#1{|{#1}\rangle}
\def\sinc{\mathop{\text{sinc}}\nolimits}
\def\cV{\mathcal{V}}
\def\cH{\mathcal{H}}
\def\cT{\mathcal{T}}
\renewcommand{\Re}{\mathop{\text{Re}}\nolimits}
\newcommand{\tr}{\mathop{\text{Tr}}\nolimits}
\newcommand{\blambda}{\boldsymbol{\lambda}}

\definecolor{dgreen}{rgb}{0,0.5,0}
\newcommand{\green}{\color{dgreen}}
\newcommand{\BLUE}[1]{\text{\color{blue} #1}}
\newcommand{\GREEN}[1]{\textbf{\color{green}#1}}
\newcommand{\REV}[1]{\textbf{\color{red}[[#1]]}}
\newcommand{\KY}[1]{\textbf{\color{dgreen}[[#1]]}}

\def\HN#1{{\color{magenta}#1}}
\def\DEL#1{{\color{red}#1}}

\title{A proposal of noise suppression for quantum annealing}

\author{Takayuki Suzuki}
\affiliation{Department of Physics, Waseda University, Tokyo 169-8555, Japan}
\author{Hiromichi Nakazato}
\affiliation{Department of Physics, Waseda University, Tokyo 169-8555, Japan}
\affiliation{Institute for Advanced Theoretical and Experimental Physics, Waseda University, Tokyo 169-8555, Japan}


\begin{abstract}
A method to suppress noise, which is one of the major obstacles to obtain an optimal solution in quantum annealers, is proposed. We generalize the conventionally used Hamiltonian, i.e., the transverse field Hamiltonian, by introducing an ancillary system, which leads to cancellation of the effect of noise on the system under consideration for some typical cases. We also confirm numerically that the method is effective for a kind of noise usually encountered in the case of flux qubit. 
\end{abstract}

\maketitle

\section{\label{sec:intro}Introduction}

Recently, quantum information technology has been actively researched and developed. Quantum annealing is one of the algorithms to solve the combinatorial optimization problem \cite{Kadowaki1998,Farhi2001}, although whether this algorithm is faster than the classical algorithms for combinatorial optimization problem still remains open. In quantum annealing, the initial state is set in the ground state of the Hamiltonian $H_0$ which is easily prepared. The final Hamiltonian $H_p$ describes the combinatorial optimization problem, which we want to solve, and the ground state of this Hamiltonian can be reached if the Hamiltonian is changed adiabatically from $H_0$ to $H_p$ thanks to the adiabatic theorem \cite{Born1928,Kato1950}. 

A device for quantum annealing composed of superconductivity qubits is produced by D-Wave Systems Inc. \cite{Harris2010,John2011}. The main weakness of the device is due to decoherence. It is known that the final state of quantum annealing in the device is not the ground state because the coherent time is shorter than the annealing time although some studies proposed to take advantage of this phenomenon for quantum annealing \cite{Dickson2013,Arceci2017,Marshall2019}. 

Even if we consider an ideal, i.e. noiseless, quantum annealer, the energy gap becomes smaller exponentially with the system size. This means that the calculation time required becomes exponentially large, which is the same as brute-force search. A previous research, however, shows that a Hamiltonian which has the term $\sigma^x\otimes \sigma^x$ 
resolves the difficulty in some models, i.e. the energy gap of the Hamiltonian becomes small not exponentially, but polynomially with the system size \cite{Seki2012}. The D-Wave device imitates transverse field Ising Hamiltonian and implementation of the term $\sigma^x\otimes \sigma^x$ is highly demanded by researchers in the field of quantum annealing \cite{Ozfidan2020}.

In this paper, we point out that a term $\sigma^x\otimes \sigma^x+\sigma^y\otimes \sigma^y$ is useful for noise suppression. Several methods to suppress the noise have already been proposed, focusing on dynamical decoupling \cite{Lidar2008}, error correction \cite{Vinci2016}, energy gap \cite{Jordan2006,Bookatz2015}, or spin-boson architecture \cite{Pino2019}. Our method is different from the previous ones. It is similar to the method proposed in \cite{Hen2016,Hen20162}, in which a different driver Hamiltonian is introduced so that the dynamics under discussion is realized in a subspace where constraints of the optimization problem are met. This method is extended to Quantum Approximate Optimization Algorithm as Quantum Alternating Operator Ansatz \cite{Hadfield2019}. We use a similar idea for noise suppression, i.e. we confine the dynamics to a subspace which is almost noise free. Remarkably, this method is successfully applicable to superconducting flux qubits, which the device of D-Wave is composed of, although the number of qubits has to be doubled.

\section{\label{sec:setup}SETUP}
In this section, we consider a closed system. In the conventional quantum annealing, the Hamiltonian is designed as a time-dependent transverse field Ising model:
\begin{align}
    H_C(t)=&-A(t)\sum_{i=1}^N \sigma_i^x+B(t)H_{C,p}^z, \label{eq:qa}\\
    H_{C,p}^z=&\sum_{i=1}^N h_i \sigma_i^z+\sum_{i=1}^N\sum_{j=i+1}^{N} J_{ij} \sigma_i^z\sigma_j^z\label{eq:pro},
\end{align}
where $A(t)$ and $B(t)$ are positive functions of time $t$. These coefficients should satisfy $A(0)\gg B(0)$ and $A(T)\ll B(T)$, where $T$ is an annealing time. We call $H_{C,p}^z$ ``problem Hamiltonian", because this term usually corresponds to the objective function of optimization problem. The initial state is set as $\otimes_{i=1}^N\ket{+}_i$ where the states are defined as $\sigma_i^x\ket{\pm}_i=\pm\ket{\pm}_i$. Because this initial state is the ground state of $H_C(0)$, the final state becomes the ground state of the problem Hamiltonian, which solves the optimization problem, if the Hamiltonian is changed adiabatically.

We propose another Hamiltonian of quantum annealing as follows:
\begin{align}
    H(t)=&A(t)\sum_{i=1}^N (c\sigma_{2i-1}^x\sigma_{2i}^x-\sigma_{2i-1}^y\sigma_{2i}^y)+B(t)H_p^z, \label{eq:nqa}\\
    H_p^z=&\sum_{i=1}^N h_{i} \sigma_{2i}^z+\sum_{i=1}^N\sum_{j=i+1}^{N} J_{ij} \sigma_{2i}^z\sigma_{2j}^z, \label{eq:prob}
\end{align}
where $c\in\mathbb{R}$ is a constant parameter. We call odd-numbered qubits ``ancilla qubits" and even-numbered qubits ``physical qubits". Note that $H_{p}^z$ is equivalent to $H_{C,p}^z$ and we also call $H_{p}^z$ ``problem Hamiltonian". This Hamiltonian \eqref{eq:nqa} does not change under the following transformation: 
\begin{align}
\begin{dcases}
    \sigma_{2i-1}^x\rightarrow -\sigma_{2i-1}^x, \ \sigma_{2i-1}^y\rightarrow -\sigma_{2i-1}^y, \ \sigma_{2i-1}^z\rightarrow \sigma_{2i-1}^z,\\
    \sigma_{2i}^x\rightarrow -\sigma_{2i}^x, \ \sigma_{2i}^y\rightarrow -\sigma_{2i}^y, \ \sigma_{2i}^z\rightarrow \sigma_{2i}^z
\end{dcases} \label{eq:trans}
\end{align}
for each $i$ in $\{1,2,\ldots,N\}$. This is due to the fact that $H(t)$ has $N$ constants of motion $\sigma_{2i-1}^z\sigma_{2i}^z \ (1\leq i\leq N)$ \cite{Grimaudo2016}. Moreover, there exists a unitary operator $C_{2i-1,2i}$ transforming $\sigma_{2i-1}^z\sigma_{2i}^z$ into $\sigma_{2i-1}^z$ because the spectrum of $\sigma_{2i-1}^z\sigma_{2i}^z$ is the same as that of $\sigma_{2i-1}^z$. The unitary operator is expressed as follows:
\begin{align}
    C_{2i-1,2i}=
    \begin{pmatrix}
    1&0&0&0\\
    0&0&0&1\\
    0&0&1&0\\
    0&1&0&0
    \end{pmatrix}
\end{align}
in the standard ordered basis
\begin{align}
    \mathcal{B}=\{\ket{00}_{2i-1,2i},\ket{01}_{2i-1,2i},\ket{10}_{2i-1,2i},\ket{11}_{2i-1,2i}\},
\end{align}
where the states are defined by $\sigma_n^z\ket{0}_n=\ket{0}_n$ and $\sigma_n^z\ket{1}_n=-\ket{1}_n$. Notice that the operator $C_{2i-1,2i}$ is nothing but a C-NOT gate $C_{2i-1,2i}=\sigma^x_{2i-1}\otimes(I_{2i}-\sigma^z_{2i})/2+I_{2i-1}\otimes(I_{2i}+\sigma^z_{2i})/2$. Let us introduce a unitary operator
\begin{align}
    \mathbb{W}\coloneqq\bigotimes_{i=1}^N C_{2i-1,2i},
\end{align}
abbreviations 
\begin{align}
    \ket{\bullet}_A\coloneqq&\ket{\bullet}_{1,3,\cdots,2N-1},\\ \ket{\bullet}_P\coloneqq&\ket{\bullet}_{2,4,\cdots,2N},
\end{align} 
and a set 
\begin{align}
    \mathcal{A}\coloneqq\{\overbrace{00\cdots00}^{N},00\cdots01,\cdots,11\cdots11\}, \label{eq:basis}
\end{align}
where the number of elements is $2^N$. Transforming $H(t)$ into $\tilde{H}(t)=\mathbb{W}^\dagger H(t) \mathbb{W}$, we get
\begin{align}
    \tilde{H}(t)=&\sum_{\blambda\in\mathcal{A}}   \ket{\blambda}_{A}\bra{\blambda}\otimes\tilde H_{\blambda}(t),\\
    \tilde H_{\blambda}(t)=&A(t)\sum_{i=1}^N (c+f(\lambda_{2i-1}))\sigma_{2i}^x \nonumber \\
    &+B(t)\left(\sum_{i=1}^N h_{i} \sigma_{2i}^z+\sum_{i=1}^N\sum_{j=i+1}^{N} J_{ij} \sigma_{2i}^z\sigma_{2j}^z\right), \label{eq:transform_qa}
\end{align}
where we have introduced a function $f(x)\coloneqq1-2x$ and $\lambda_{2i-1}$ is the $i$-th element of $\blambda$, i.e., $\blambda=(\lambda_1,\lambda_3,\cdots,\lambda_{2N-1})$. We can see that the Hamiltonian $\tilde H_{\blambda}(t)$ is the same as $H_C(t)$ in \eqref{eq:qa} up to the coefficient of transverse field. 

Let us consider the dynamics of conventional quantum annealing using $\tilde H_{\blambda}(t)$ in \eqref{eq:transform_qa}. We define the time evolution operator corresponding to $\tilde{H}_{\blambda}(t)$ as $\tilde U_{\blambda} (t)$. We also express the basis for subspace of physical qubits as follows:
\begin{align}
     \left\{\bigotimes_{i=1}^N \ket{s_{n,i}}_{2i}\right\}_{n=1}^{2^N}
     =&\{\ket{00\cdots0}_{P},\ket{00\cdots1}_{P},\cdots,\ket{11\cdots1}_{P}\}.
\end{align}
If $c+f(\lambda_{2i-1})$ in \eqref{eq:transform_qa} is negative for all $i$, the ground state which is chosen as initial state is $\bigotimes_{i=1}^N\ket{+}_{2i}$. The state at $t$ is expressed as $\tilde U_{\vec{1}}(t)\bigotimes_{i=1}^N\ket{+}_{2i}$, where $\blambda=\vec{1}$ means $\blambda=(1,1,\cdots,1)$, . We define $a_n(t)$ as the coefficient of the state when the state is expanded with the basis in \eqref{eq:basis}:
\begin{align}
    \tilde U_{\vec{1}}(t)\bigotimes_{i=1}^N\ket{+}_{2i} =\sum_{n=1}^{2^N}\left(a_n(t) \bigotimes_{i=1}^N \ket{s_{n,i}}_{2i}\right). \label{eq:timeevolution_usual}
\end{align}

Next, we calculate the time evolution in the original space. The Schr\"odinger equation
\begin{align}
    i\frac{d}{dt} U(t)=&H(t) U(t)
\end{align}
is transformed to
\begin{align}
    i\frac{d}{dt}\mathbb{W}^\dagger U(t)\mathbb{W}=&\tilde H(t) \mathbb{W}^\dagger U(t)\mathbb{W}\nonumber \\
    =&\sum_{\blambda\in\mathcal{A}} \ket{\blambda}_{A}\bra{\blambda}\otimes \tilde H_{\blambda}(t)  \mathbb{W}^\dagger U(t)\mathbb{W}. \label{eq:proof}
\end{align}
The Hamiltonian in \eqref{eq:proof} indicates that the subdynamics of the physical qubits connected to each ancilla qubit does not interfere each other. In other words, the time evolution operator can be expressed as follows:
\begin{align}
    \mathbb{W}^\dagger U(t)\mathbb{W}= \sum_{\blambda\in \mathcal{A}}\ket{\blambda}_{A}\bra{\blambda}\otimes\tilde U_{\blambda}(t)  . \label{eq:tildeU}
\end{align}
We can easily confirm this equality by substituting \eqref{eq:tildeU} for \eqref{eq:proof}. Therefore, the time evolution operator corresponding to $H(t)$ in \eqref{eq:nqa} is represented as
\begin{align}
    U(t)=\mathbb{W} \left(\sum_{\blambda\in \mathcal{A}}\ket{\blambda}_{A}\bra{\blambda}\otimes \tilde U_{\blambda}(t)  \right)\mathbb{W}^\dagger.
\end{align}
If the initial state is
\begin{align}
    \ket{\psi_I}=\bigotimes_{i=1}^N\frac{1}{\sqrt{2}}(\ket{01}_{2i-1,2i}+\ket{10}_{2i-1,2i}),
\end{align}
then the state at $t$ is calculated as follows:
\begin{align}
    \ket{\psi(t)}=&U(t)\bigotimes_{i=1}^N\frac{1}{\sqrt{2}}(\ket{01}_{2i-1,2i}+\ket{10}_{2i-1,2i})\nonumber \\
    =&\mathbb{W} \left(\sum_{\blambda\in \mathcal{A}}\ket{\blambda}_{A}\bra{\blambda}\otimes  \tilde U_{\blambda}(t) \right)\bigotimes_{i=1}^N\ket{1}_{2i-1}\ket{+}_{2i}\nonumber \\
    =&\mathbb{W} \left(\ket{\vec{1}}_{A}\otimes\tilde U_{\vec{1}}(t)\bigotimes_{i=1}^N \ket{+}_{2i}\right) \nonumber \\
    =&\mathbb{W} \left(\ket{\vec{1}}_{A} \otimes\sum_{n=1}^{2^N}\left(a_n(t) \bigotimes_{i=1}^N \ket{s_{n,i}}_{2i}\right)\right) \nonumber \\
    =&\sum_{n=1}^{2^N}\left(a_n(t)  \bigotimes_{i=1}^N\ket{\bar{s}_{n,i}s_{n,i}}_{2i-1,2i}\right),
\end{align}
where we define $\ket{\bar{0}}=\ket{1}$ and $\ket{\bar{1}}=\ket{0}$. If only the physical qubits are measured, the probability that we get a state $\otimes_{i=1}^N\ket{s_{m,i}}_{2i}$ is $|a_m(t)|^2$. This is the same as the usual quantum annealing, as can be seen from \eqref{eq:timeevolution_usual}. We should set $c<0$ so that $\ket{\psi_I}$ is the ground state of the initial Hamiltonian $H(0)$.

\section{\label{sec:noisesup}Noise Suppression}
In this section, we discuss how and when noise can be suppressed using the proposed Hamiltonian \eqref{eq:nqa} when the system is open and exposed to the noise. In the conventional quantum annealing, such an open dynamics is described by the following total Hamiltonian \cite{Albash2012}:
\begin{align}
    H_{tot,C}(t)=&H_C(t)+\sum_{k=1}^\infty \omega_k b_k^\dagger b_k \nonumber \\
    &+\sum_{i=1}^{N}\sum_{k=1}^\infty\left(g^z_{i,k} \sigma_{i}^z+g_{i,k}^x\sigma_i^x\right)\otimes\left(b_k^\dagger+b_k\right),
    \label{eq:conventional_total_hamiltonian}
\end{align}
where $b_k$ and $b_k^\dagger$ are bosonic annihilation and creation operators satisfying the standard commutation relations $[b_k,b_{k'}^\dagger]=\delta_{kk'}$ etc., and consist of a bosonic reservoir. The interaction between the $i$-th spin of the system and the mode $k$ of the reservoir is assumed to take place through the longitudinal and transversal couplings with the form factors $g^z_{i,k}$ and $g^x_{i,k}$, respectively. It has been argued \cite{Albash2012} that the effect of noise introduced by such couplings brings about considerable degradations in the quantum annealing process. In this paper, the conventional system Hamiltonian $H_C(t)$ is replaced with our proposal $H(t)$ in \eqref{eq:nqa} and we discuss the open dynamics described by the total Hamiltonian
\begin{align}
    H_{tot}(t)=&H(t)+\sum_{k=1}^\infty \omega_k b_k^\dagger b_k \nonumber \\
    &+\sum_{i=1}^{2N}\sum_{k=1}^\infty\left(g^z_{i,k} \sigma_{i}^z+g_{i,k}^x\sigma_i^x\right)\otimes\left(b_k^\dagger+b_k\right).
    \label{eq:proposed_total_hamiltonian}
\end{align}
Observe that our system, composed of physical (even-numbered) and ancillary (odd-numbered) spins, is interacting with the common reservoir.

\subsection{Effect of longitudinal ($\sigma^z$) coupling}
In this subsection, we consider the case where the transversal couplings are negligible $|g^z_{i,k}|\gg|g^x_{i,k}|\sim0$ $\forall i,k$ because it is recognized that this inequality holds in superconducting
flux qubits \cite{Harris2010,John2011,Marshall2019,Boixo2016}. In this case, the total Hamiltonian can be approximated as follows:
\begin{align}
    H_{tot}(t)=&H(t)+\sum_{k=1}^\infty \omega_k b_k^\dagger b_k \nonumber \\
    &+\sum_{i=1}^{2N}\sum_{k=1}^\infty g^z_{i,k} \sigma_{i}^z\otimes\left(b_k^\dagger+b_k\right).
\end{align}
This total Hamiltonian does not change under the transformation \eqref{eq:trans} and can be transformed in the same way. Transforming $H_{tot}(t)$ into $\tilde{H}_{tot}(t)=\mathbb{W}^\dagger H_{tot}(t) \mathbb{W}$, we get
\begin{align}
    \tilde{H}_{tot}(t)=&\sum_{\blambda\in\mathcal{A}} \ket{\blambda}_{A}\bra{\blambda}\otimes  \tilde H_{tot,\blambda}(t),\\
    \tilde H_{tot,\blambda}(t)=&\tilde H_{\blambda}(t) +\sum_{k=1}^\infty \omega_k b_k^\dagger b_k\nonumber \\ +\sum_{i=1}^N\sum_{k=1}^\infty&\left(f(\lambda_{2i-1})g_{2i-1,k}^z+g_{2i,k}^z\right)\sigma_{2i}^z\otimes\left(b_k^\dagger+b_k\right).
\end{align}
This indicates that the noise does not affect the system dynamics, i.e., the system decouples from the noisy environment, if $\lambda_{2i-1}=1$ and $g_{2i-1,k}^z=g_{2i,k}^z$ hold for arbitrary $i$ and $k$. In this case, the time evolution of the spin system is the same as that of a closed system. Even if such exact equalities do not hold, when the differences $|g_{2i-1,k}^z-g_{2i,k}^z|$ are much smaller than $A(t), B(t)$, and $\omega_k$ for arbitrary $i$ and $k$, the spin system and boson bath can be regarded as approximately decoupled. It is remarkable that the effect of noise caused through the longitudinal couplings can be completely (or at least almost) canceled in the present protocol proposed here.   

\subsection{Effect of both the longitudinal and transversal couplings: Numerical simulations}
In this subsection, we numerically examine the dynamics in \eqref{eq:proposed_total_hamiltonian} when the transversal couplings are not negligible. In the numerical calculation, we simulate the dynamics in terms of (nearly) adiabatic quantum master equation \cite{Albash2012,yamaguchi17}. According to the microscopic derivation in \cite{Albash2012}, the master equation derived reads as follows (we assume an Ohmic bath and ignore the Lamb shift for simplicity):
\begin{align}
    \dot\rho(t)=&-i[H(t),\rho(t)]\nonumber \\
    &+\sum_{\alpha=x,z}\sum_{\omega}\sum_{i,j=1}^{2N} g_{i,|\omega|}^{\alpha}g_{j,|\omega|}^{\alpha}\gamma(\omega)\nonumber\\
    &\hspace{5mm} \times \biggl[L_{\alpha,j ,\omega}(t)\rho(t)L^\dagger_{\alpha,i,\omega}(t)\nonumber \\
    &\hspace{10mm} -\frac{1}{2}\biggl\{L^\dagger_{\alpha,i,\omega}(t)L_{\alpha,j,\omega}(t),\rho(t)\biggr\}\biggr], \label{eq:master}\\
    L_{\alpha,i,\omega}(t)=&\sum_{\omega_{ba}(t)=\omega}\ket{E_a(t)}\bra{E_a(t)}\sigma_i^\alpha\ket{E_b(t)}\bra{E_b(t)} ,\label{eq:lind}\\
    \gamma(\omega)=&\frac{\eta |\omega| e^{-\frac{|\omega|}{\omega_c}}}{1-e^{-\beta |\omega|}}(\Theta(\omega)+\Theta(-\omega)e^{-\beta|\omega|})\label{eq:realx},
\end{align}
where $\beta$ is the inverse temperature of the bosonic bath, $\omega_c$ is a high-frequency cutoff and $\eta$ is a positive constant. We defined the instantaneous eigenstates and eignevalues of $H(t)$ as $\ket{E_a(t)}$ and $E_a(t)$, the energy gaps as $\omega_{ba}(t)=E_{b}(t)-E_{a}(t)$, and the Heaviside step function as $\Theta(\omega)$. In (\ref{eq:realx}), only one of the step functions is supposed to be non-zero at $\omega=0$. We note that this master equation is derived under the assumption that the Hamiltonian of the system is adiabatic \cite{Albash2012} or nearly adiabatic \cite{yamaguchi17}.

\begin{figure}[h]
    \begin{minipage}[b]{0.475\linewidth}
    \begin{center}
    \includegraphics[width=37.5mm]{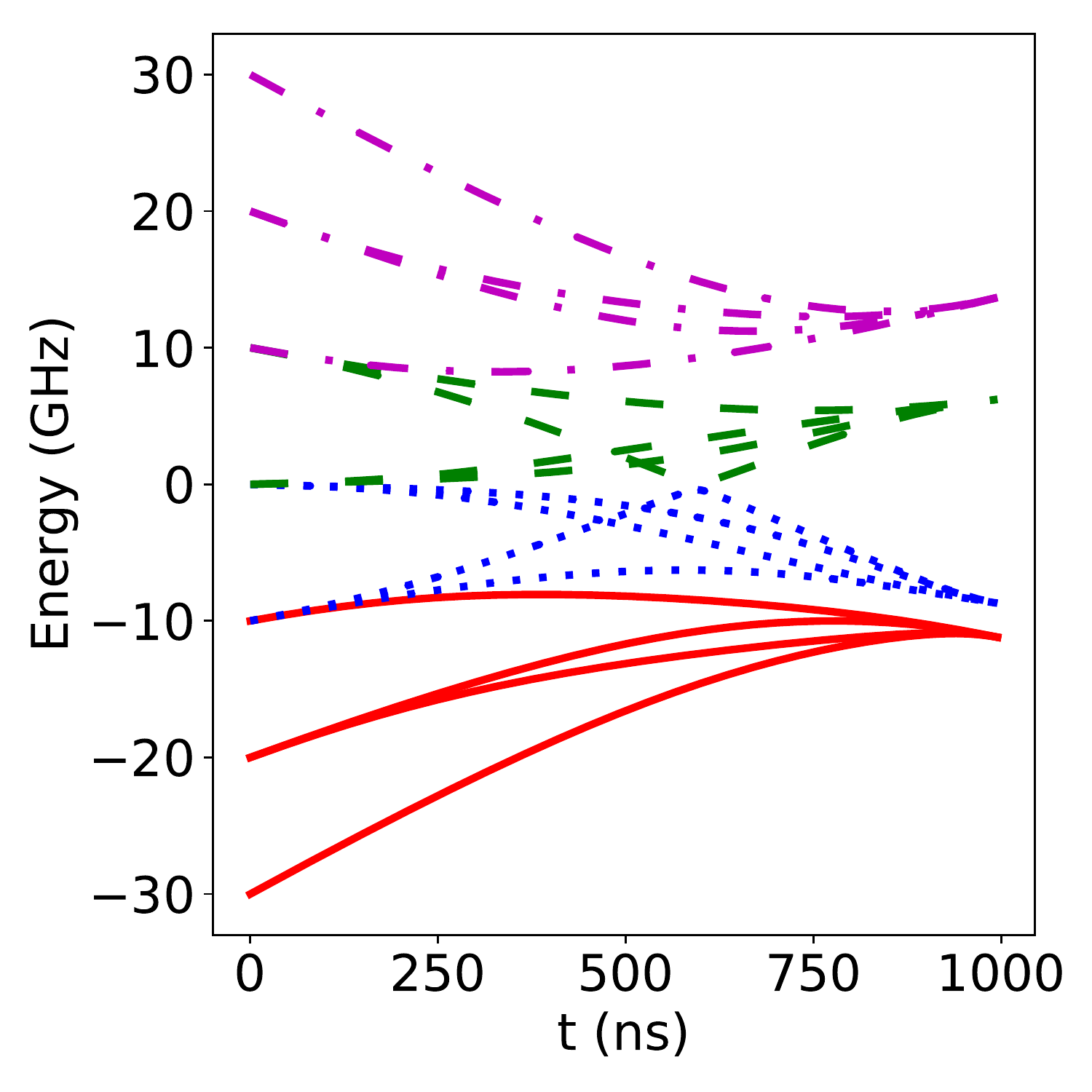}
    \end{center}
    \end{minipage}
    \begin{minipage}[b]{0.475\linewidth}
    \begin{center}
    \includegraphics[width=37.5mm]{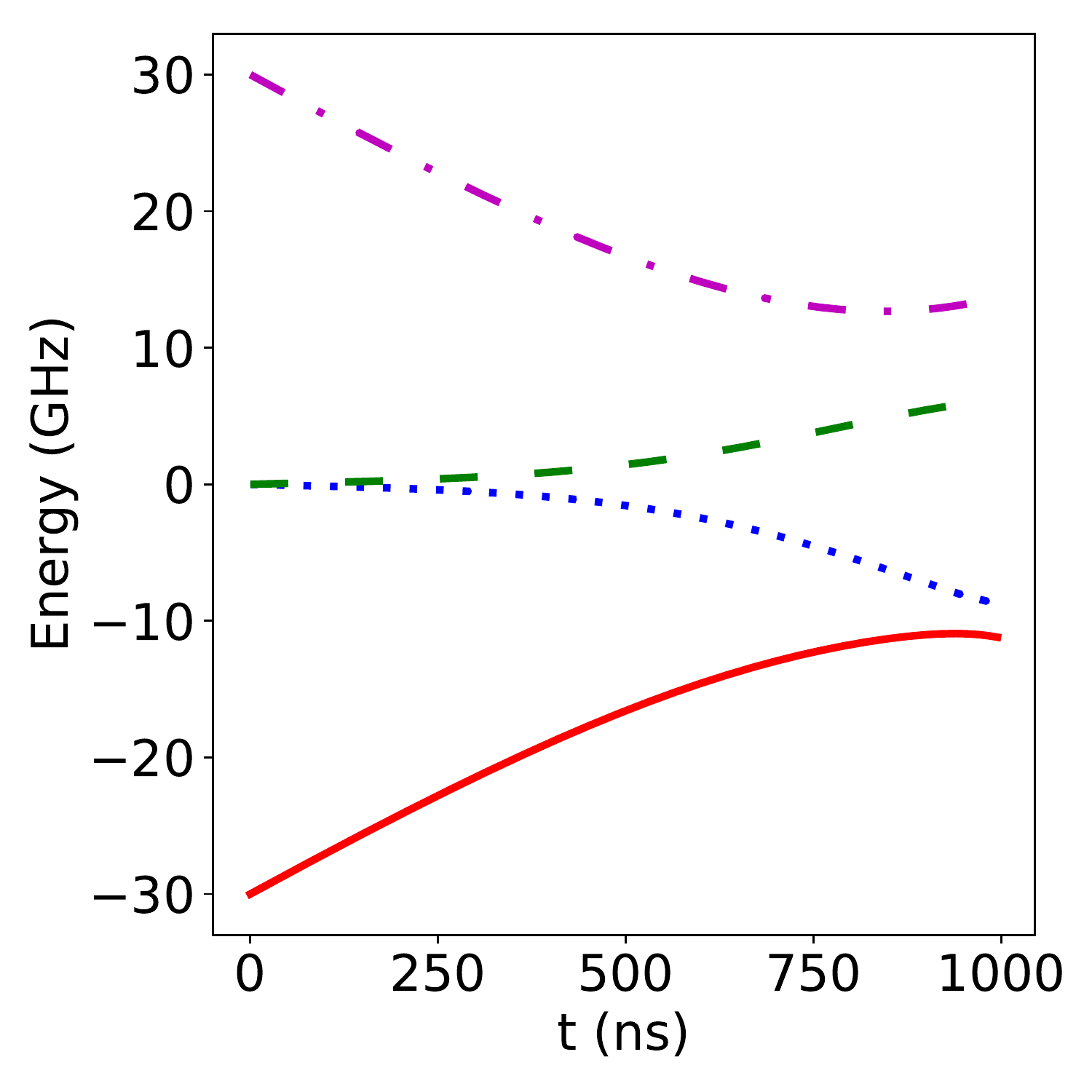}
    \end{center}
    \end{minipage}
    \caption{Time dependence of energy eigenvalues of $H(t)$ (left) and $\tilde{H}_{\vec{1}}(t)$ (right) with $c=-1/2$. The problem Hamiltonian is given in \eqref{eq:prob} with parameters \eqref{eq:coef} and we set $A(t)=at/T, \ B(t)=a-A(t)$, and $a=10$GHz.} 
\label{fig:ene}
\end{figure}

We investigate the effect of $g^x_{i,k}$ in the case of $N=2$, i.e., the number of qubits is $4$. We set $A(t)=at/T, \ B(t)=a-A(t)$, and $a=10$GHz (in units such that $\hbar=1$). Note that this energy scale is the same as the energy scale of D-Wave 2000Q quantum annealer \cite{Marshall2019}. 
For concreteness, we consider the problem Hamiltonian \eqref{eq:prob} with the following parameters:
\begin{align}
    h_2=1, \ h_4=\frac{1}{4}, \ J_{24}=\frac{1}{8}. \label{eq:coef}
\end{align}
The ground state at $t=T$ is $\ket{E_0(T)}=\ket{11}_{24}\eqqcolon\ket{g}$. In the following numerical calculation, we set $T=1000$ns and consider a uniform noise, i.e. $g_{i,k}^x=g_x$ and $g_{i,k}^z=g_z$. We also set $1/\beta=1.57$GHz $\simeq 12$mK, $\eta=0.2($GHz$)^{-2}$, and $\omega_c=8\pi$GHz in accordance with \cite{Albash2012}. Time dependence of the energy spectrum is shown in FIG. \ref{fig:ene}.

\begin{figure}[h]
    \begin{minipage}[b]{0.42\linewidth}
    \begin{center}
    \includegraphics[width=52mm]{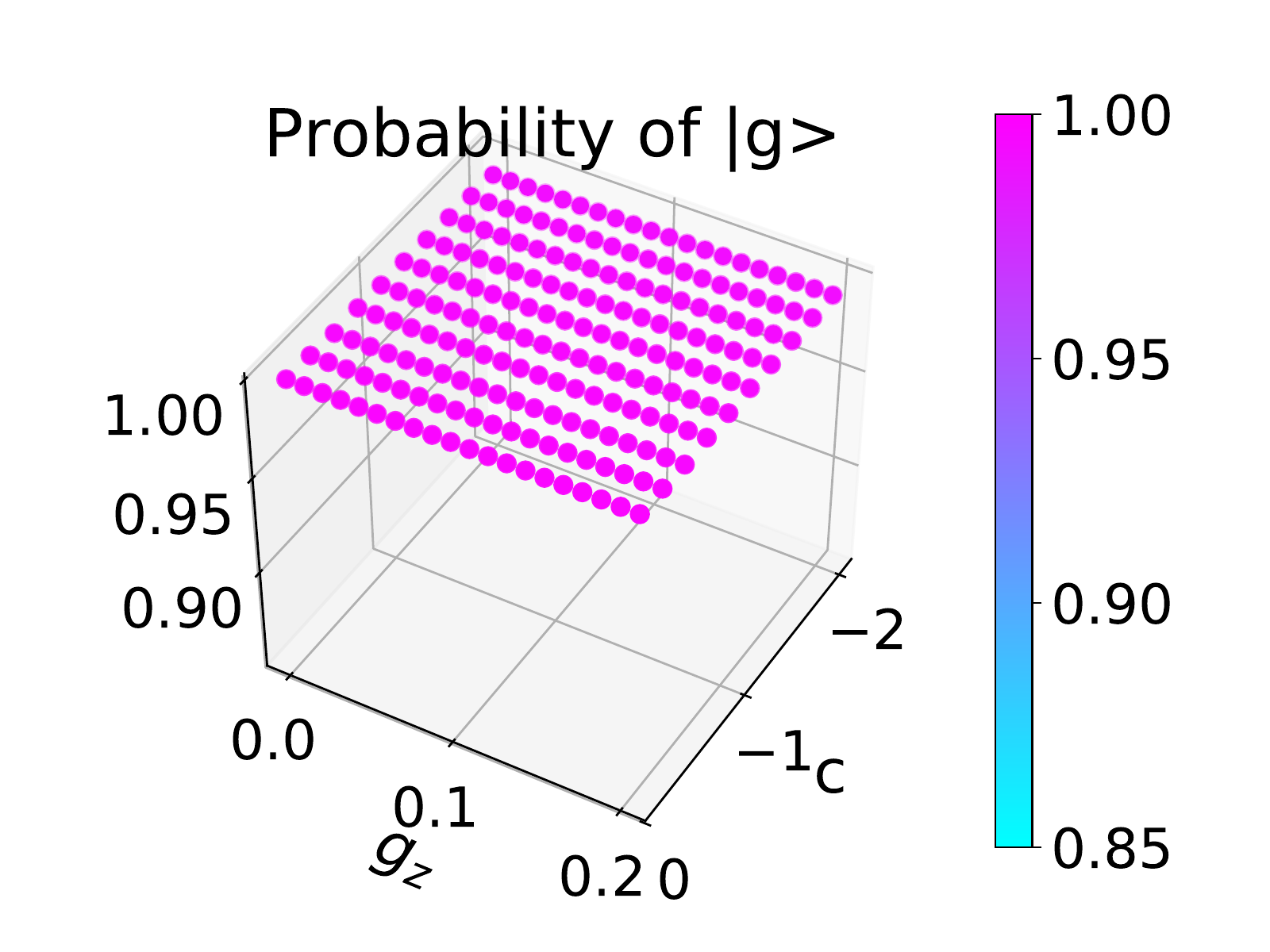}
    \end{center}
    \end{minipage}
    \begin{minipage}[b]{0.56\linewidth}
    \begin{center}
    \includegraphics[width=52mm]{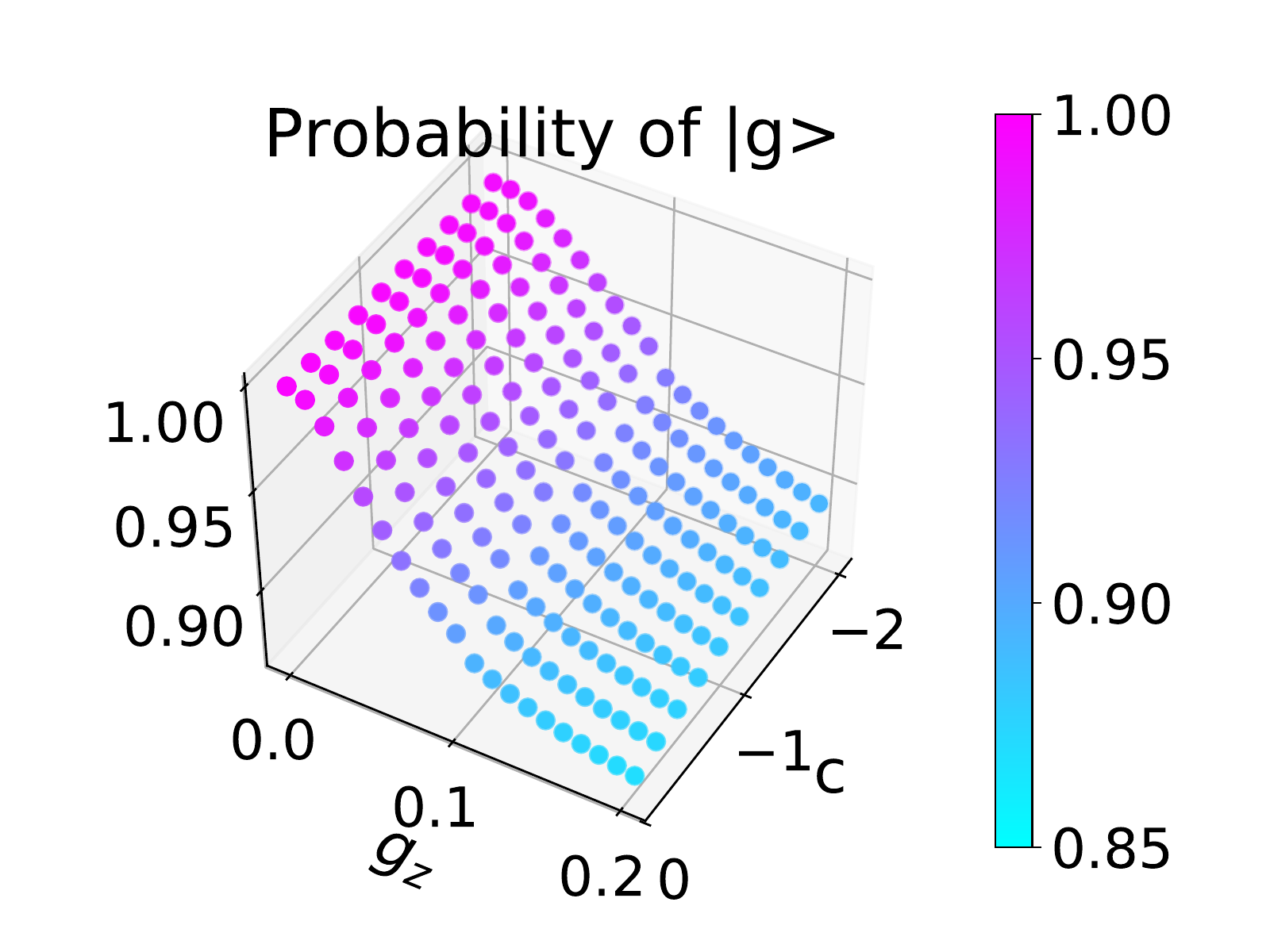}
    \end{center}
    \end{minipage}\\
    \begin{minipage}[b]{0.98\linewidth}
    \begin{center}
    \includegraphics[width=57.5mm]{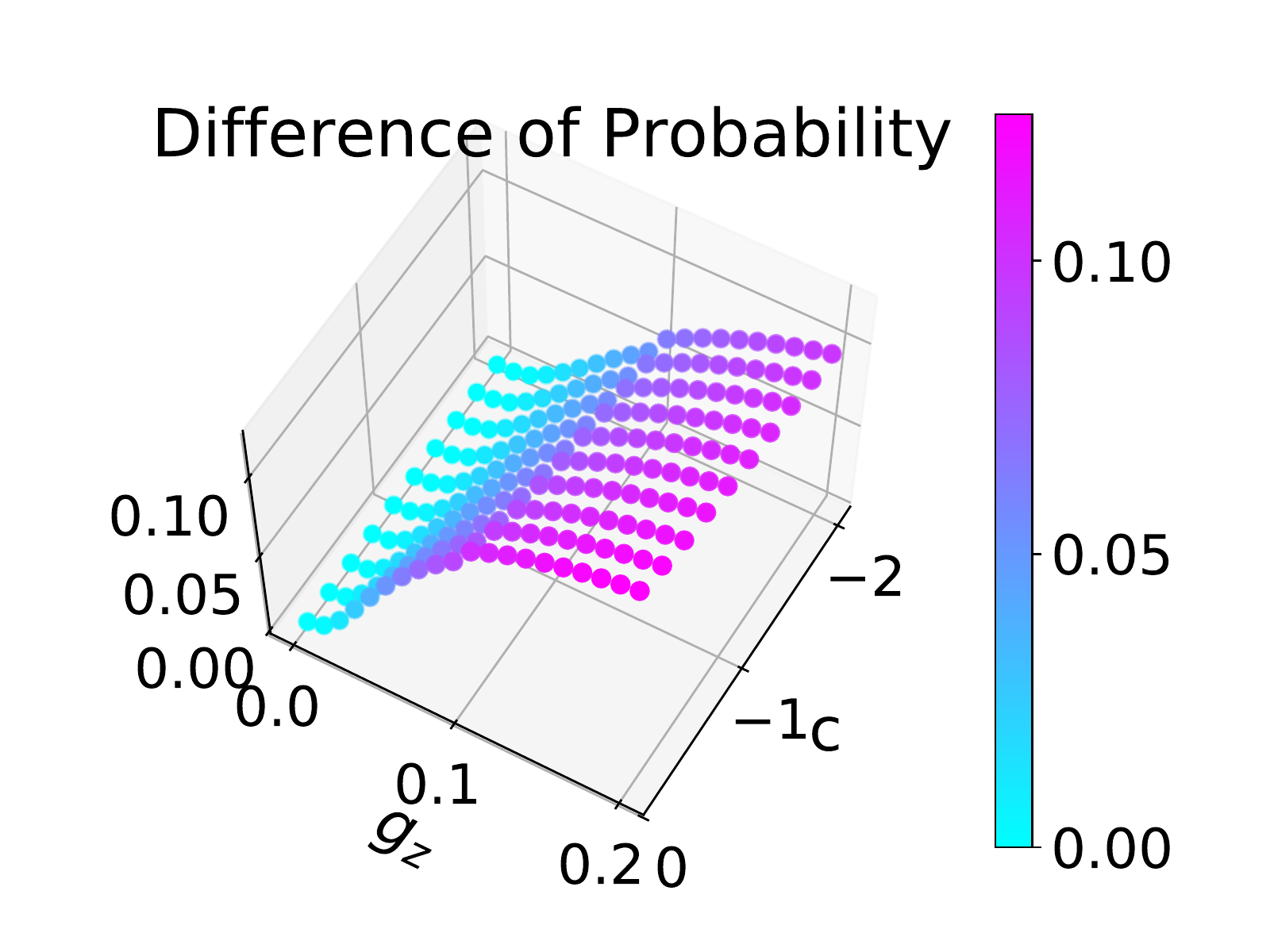}
    \end{center}
    \end{minipage}
    \caption{Probabilities to measure the ground state at $t=T$, according to the present scheme $H_{tot}(t)$ (left) and to the standard one $H_{tot,C}(t)$ (right) with $g_{i,k}^x=0$ and $g^z_{i,k}=g_z$. Their difference is shown at (bottom). We set $1/\beta=1.57$GHz $\simeq 12$mK, $\eta=0.2($GHz$)^{-2}$, and $\omega_c=8\pi$GHz.}
\label{fig:c_gz}
\end{figure}

First, consider the case $g_x=0$. Figure \ref{fig:c_gz} shows the probability that the ground state is obtained when measured at $t=T$. As we explained before, noises are cancelled completely in the proposed Hamiltonian \eqref{eq:proposed_total_hamiltonian}. We note that the quantum adiabatic master equation \eqref{eq:master} is valid in this case because the dynamics is adiabatic. This is why the probability to measure the ground state is almost $1$ in the present scheme.

\begin{figure}[H]
    \begin{minipage}[b]{0.42\linewidth}
    \begin{center}
    \includegraphics[width=52mm]{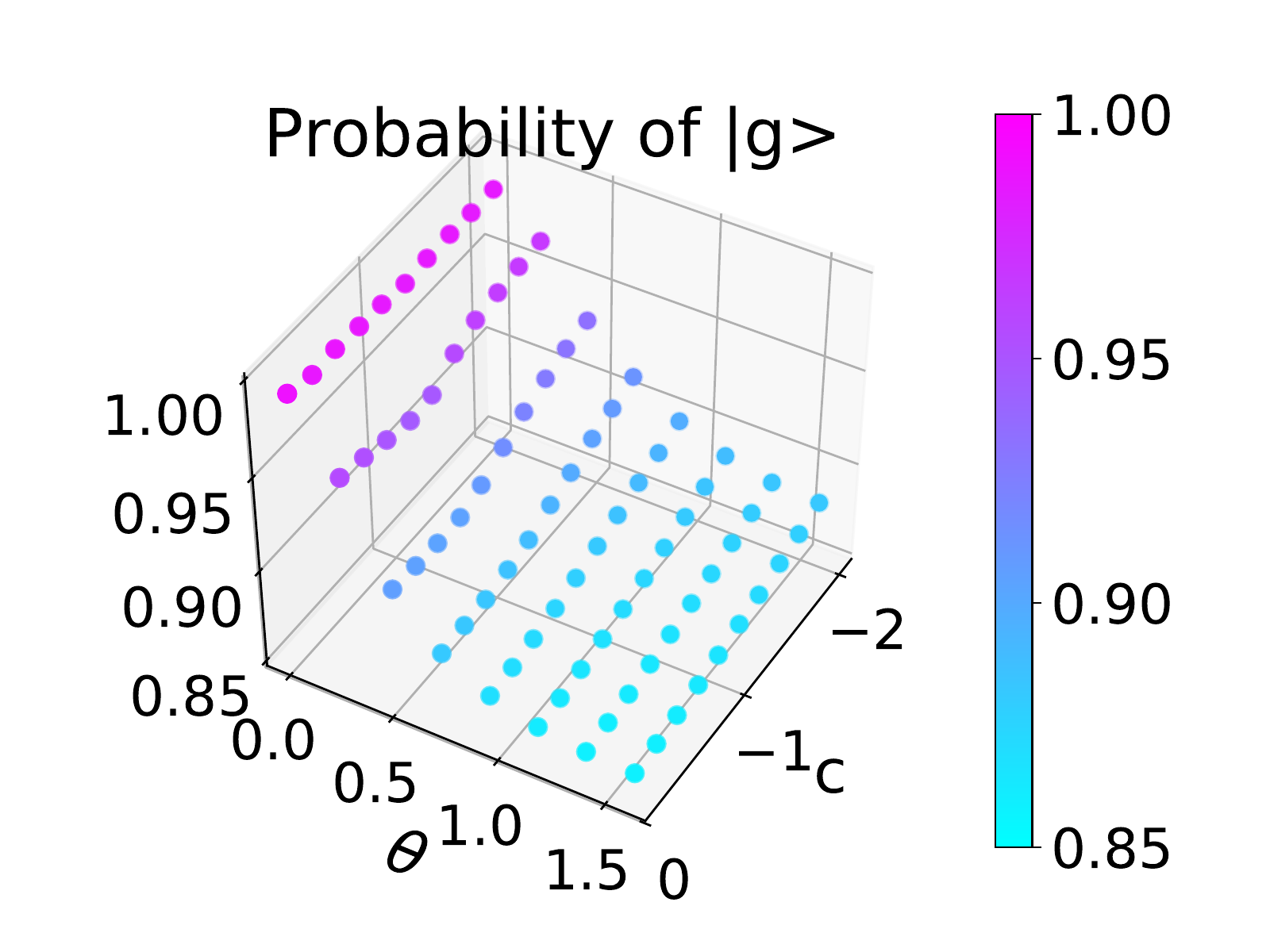}
    \end{center}
    \end{minipage}
    \begin{minipage}[b]{0.56\linewidth}
    \begin{center}
    \includegraphics[width=52mm]{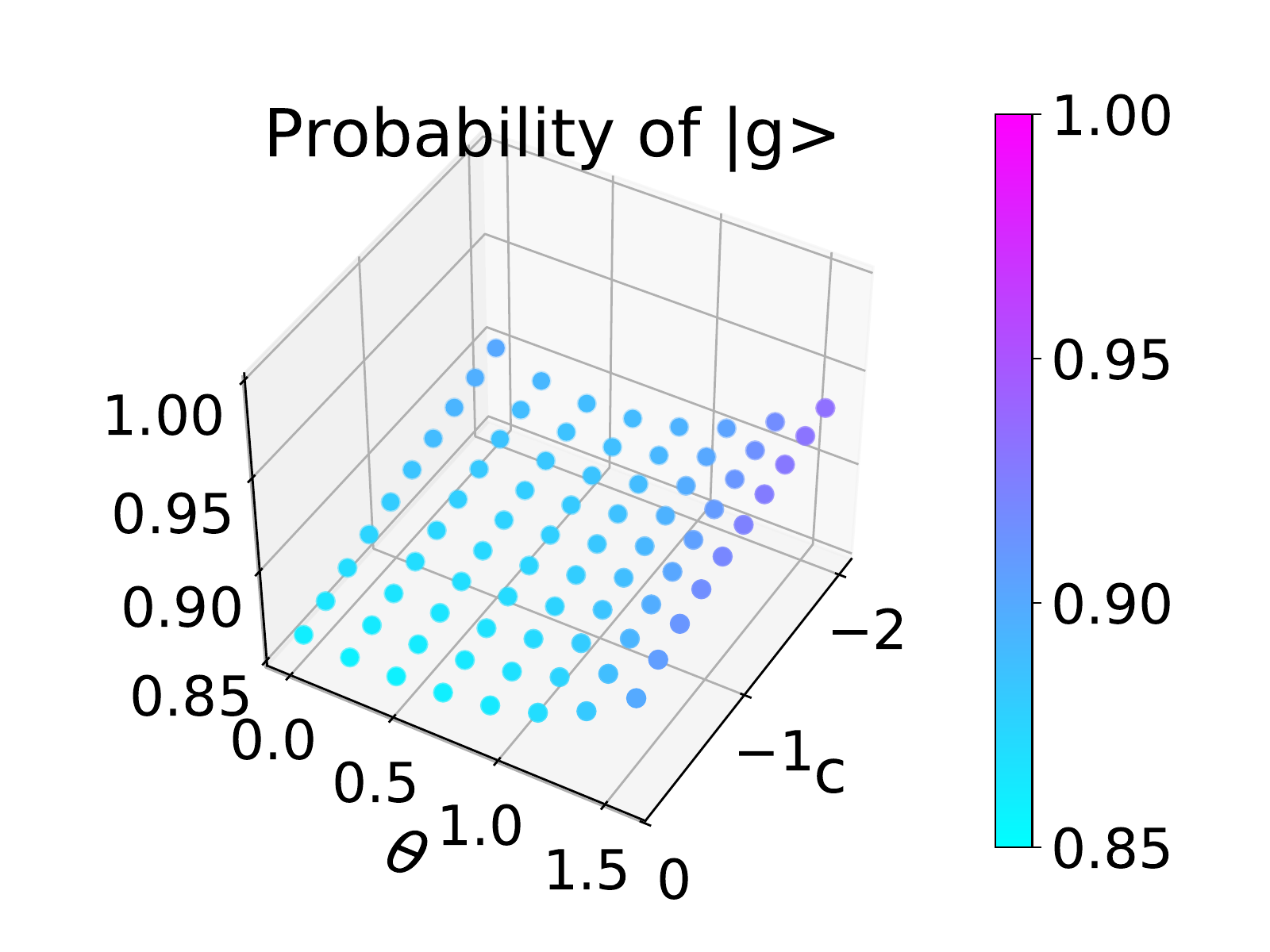}
    \end{center}
    \end{minipage}\\
    \begin{minipage}[b]{0.98\linewidth}
    \begin{center}
    \includegraphics[width=57.5mm]{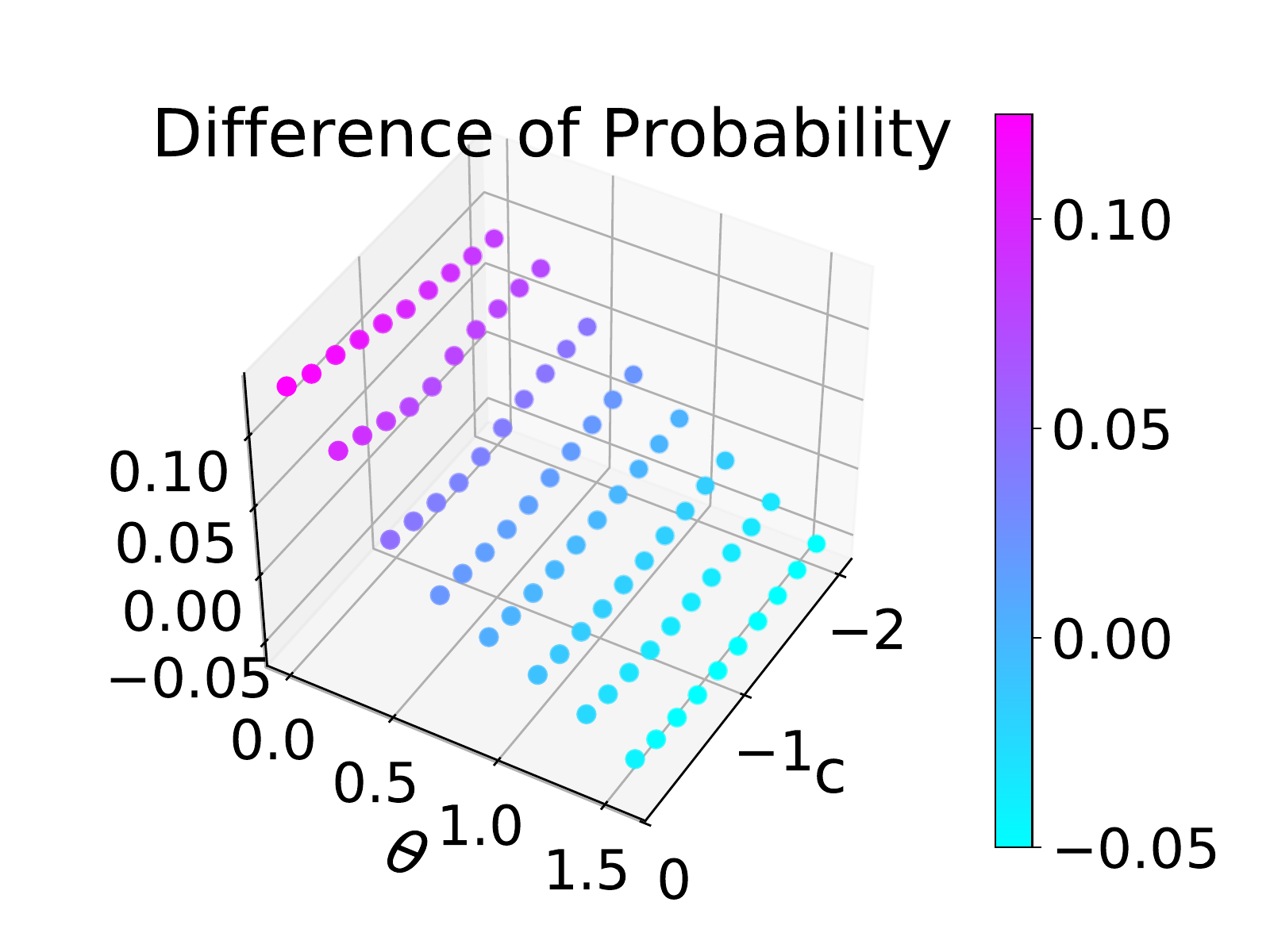}
    \end{center}
    \end{minipage}
    \caption{Probabilities to measure the ground state at $t=T$, according to the present scheme $H_{tot}(t)$ (left) and to the standard one $H_{tot,C}(t)$ (right) with $g_{i,k}^x=g\sin\theta$, $g^z_{i,k}=g\cos\theta$, and $g=0.1$GHz. Their difference is shown at (bottom). We set $1/\beta=1.57$GHz $\simeq 12$mK, $\eta=0.2($GHz$)^{-2}$, and $\omega_c=8\pi$GHz.}
\label{fig:c_theta}
\end{figure}

\begin{figure}[H]
    \begin{minipage}[b]{0.42\linewidth}
    \begin{center}
    \includegraphics[width=52mm]{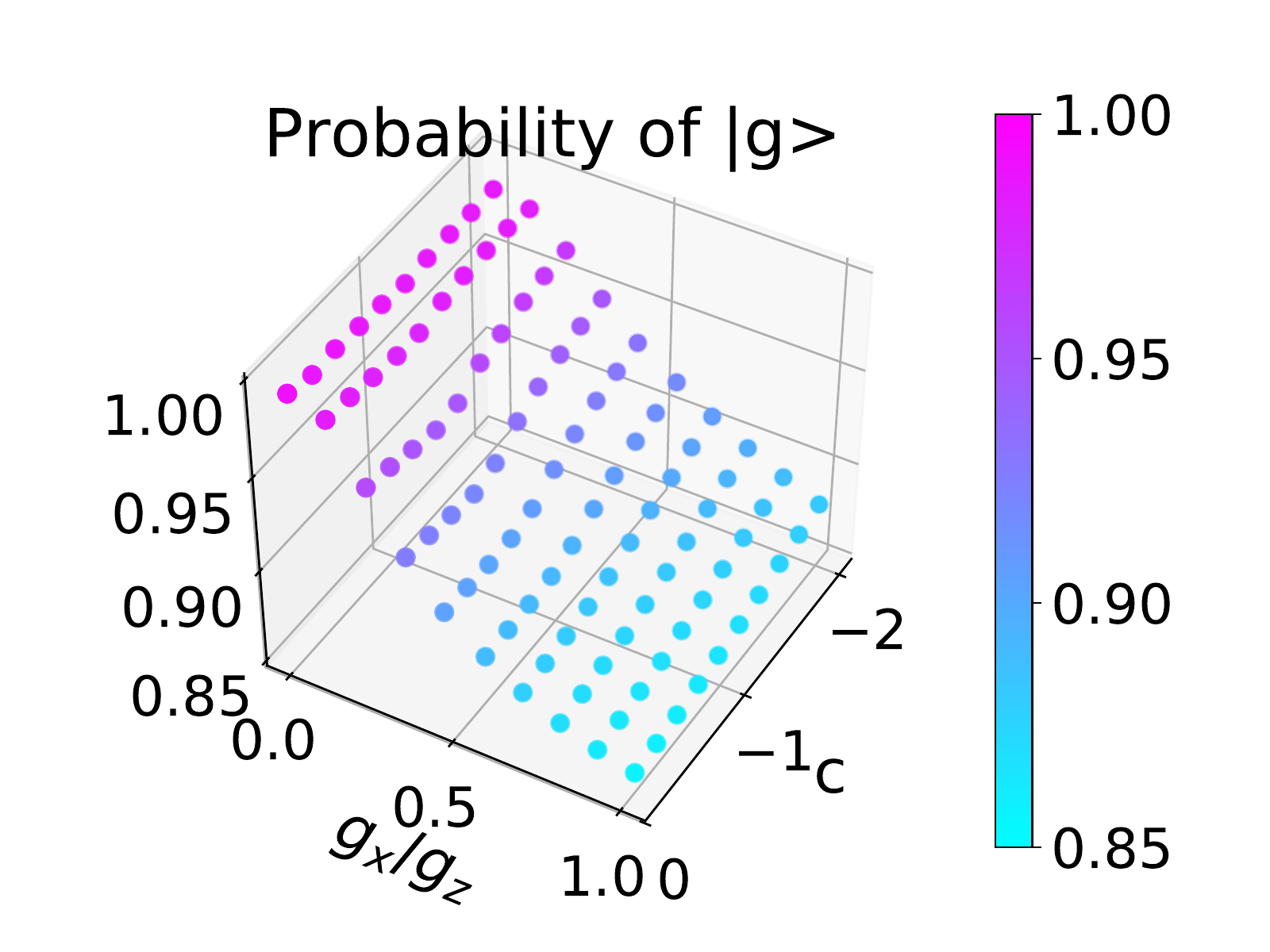}
    \end{center}
    \end{minipage}
    \begin{minipage}[b]{0.56\linewidth}
    \begin{center}
    \includegraphics[width=52mm]{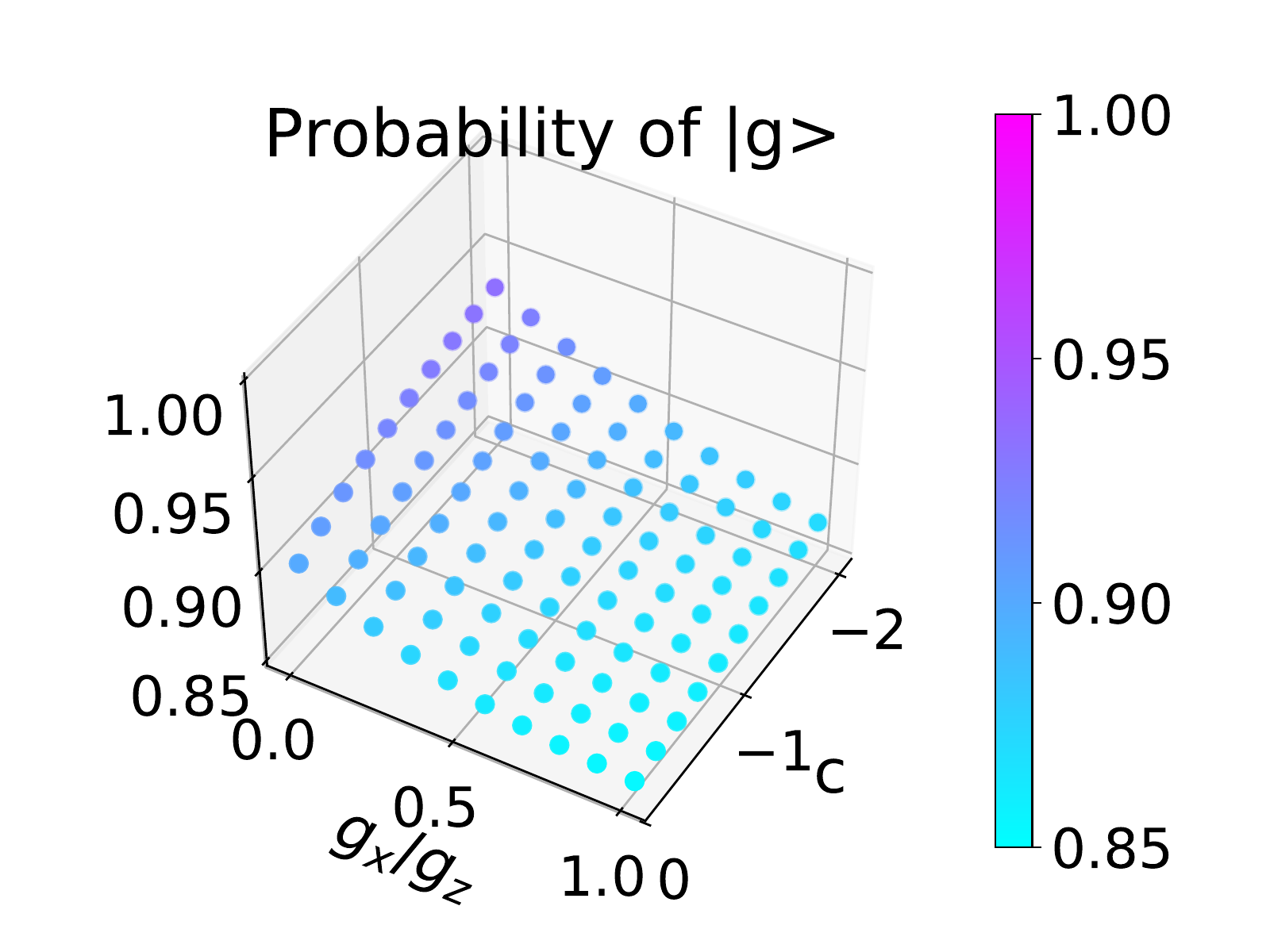}
    \end{center}
    \end{minipage}\\
    \begin{minipage}[b]{0.98\linewidth}
    \begin{center}
    \includegraphics[width=57.5mm]{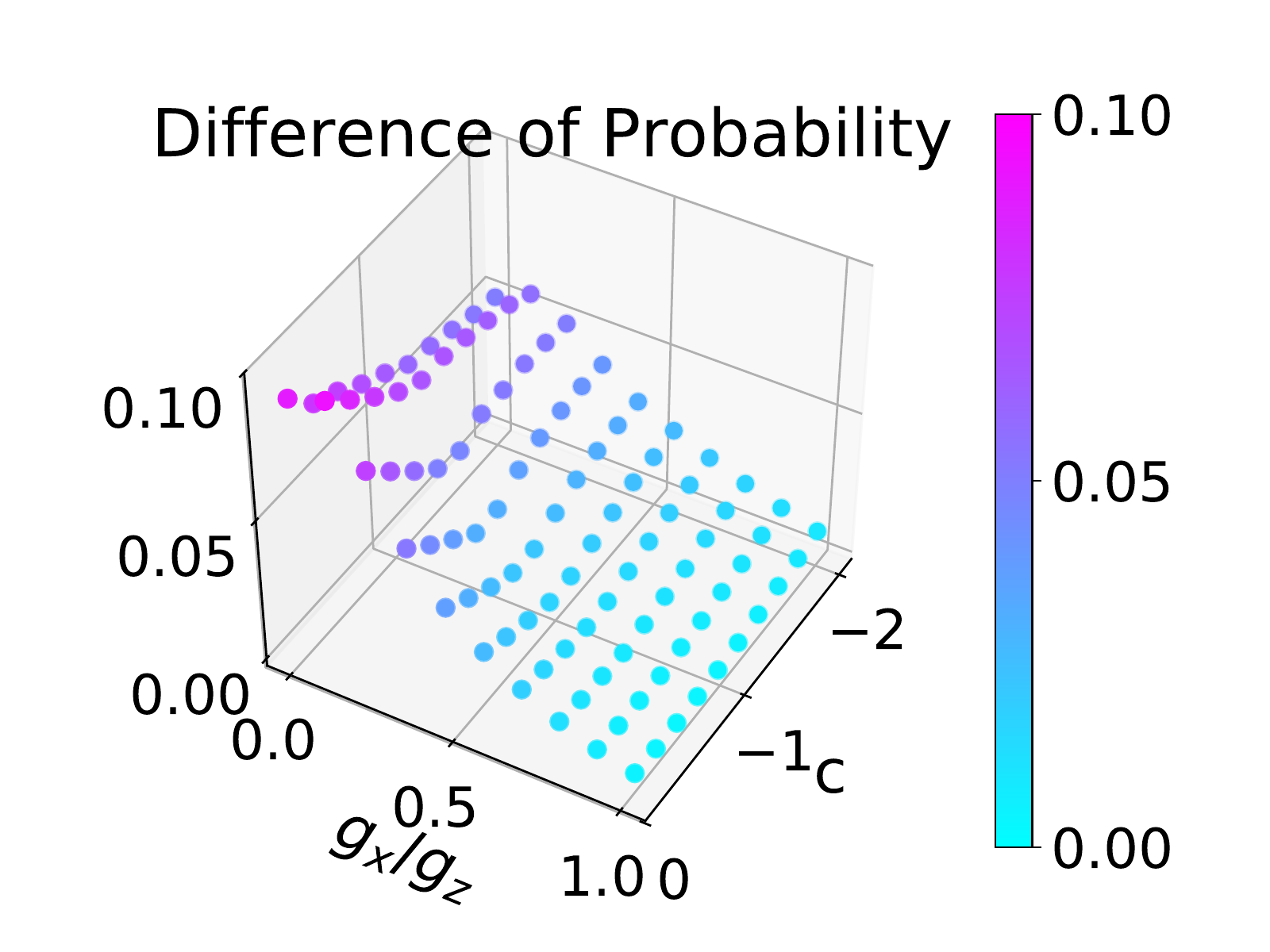}
    \end{center}
    \end{minipage}
    \caption{Probabilities to measure the ground state at $t=T$, according to the present scheme $H_{tot}(t)$ (left) and to the standard one $H_{tot,C}(t)$ (right) with $g_{i,k}^x=g_x$ and $g^z_{i,k}=g_z=0.1$GHz. Their difference is shown at (bottom). We set $1/\beta=1.57$GHz $\simeq 12$mK, $\eta=0.2($GHz$)^{-2}$, and $\omega_c=8\pi$GHz.}
\label{fig:c_gx}
\end{figure}

Next, we consider the case $g_x\neq0$. We consider two patterns: 1) the strength of the noise $g=\sqrt{(g_x)^2+(g_z)^2}$ is fixed and the relative weight measured by $\tan\theta=g_x/g_z$ is varied and 2) only $g_x$ is varied with $g_z$ kept fixed. The results of $g=0.1$GHz for case 1) is shown in FIG.\ref{fig:c_theta} and the result of $g_z=0.1$GHz for case 2) is shown in FIG.\ref{fig:c_gx}. From these figures, we understand that in the present scheme, the noise can be almost cancelled when $g_x\lesssim 0.1g_z$. We note that proposed method would lose its superiority over the conventional one only when $\theta\simeq \pi/2$.

\section{SUMMARY}
We have reported a method to suppress noise in quantum annealing. In this proposal, we introduce a driver Hamiltonian different from conventional one to make a subspace where there is almost no noise. In the example, we showed the superiority of the proposed method in the problem where the number of variables is $2$. In this case, the probability to measure the ground state at $t=T$ is not very low even in the conventional method because the energy gaps are large enough. If the number of qubits becomes larger, however, the superiority of the proposed method will become more dramatic because the energy gaps become smaller in general, which causes no problem in the current scheme that is immune to decoherence caused by longitudinal noise. This idea is also valid for adiabatic quantum computer if the type of noise is the same. 

We emphasize that the proposed method is completely new and is different from the idea of decoherence-free subspace \cite{Duan1998,Zanardi1997,Lidar1998,Zanardi1998}. There are two main differences. First, in the idea of decoherence-free subspace, the state evolves according to the Schr\"odinger equation with an effective Hamiltonian which is different from the original system Hamiltonian. In contrast, the state evolves under the system Hamiltonian itself in this proposal. Second, in the idea of decoherence-free subspace, the state should be in a space that is spanned by the eigenstates of Lindblad operators. If Lindblad operators depend on time, it is usually difficult to hold the condition \cite{Wu2012}. In our proposal, however, such a condition is not required because the coupling cancels if it is almost longitudinal. This method could be utilized to suppress noise in future quantum devices.

\section*{ACKNOWLEDGMENTS}
H. N. is partly supported by Waseda University Grant for Special Research Projects 2020-C272. T. S. is partly supported by Top Global University Project, Waseda University.

\appendix
\section{Eigenvalues/ Eigenstates}
We prove that the eigenvalues of $\tilde{H}_{\blambda}(t)$ are included in those of $H(t)$. The eigenequation of $\tilde{H}_{\blambda}(t)$ is as follows:
\begin{align}
    \tilde{H}_{\blambda}(t)\ket{\tilde{\varepsilon}_{n,\blambda}(t)}_P=&\tilde{\varepsilon}_{n,\blambda}(t)\ket{\tilde{\varepsilon}_{n,\blambda}(t)}_P.
\end{align}
This equation can be transformed as follows:
\begin{align}
    \sum_{\blambda\in\mathcal{A}}\left(\ket{\blambda}_A\bra{\blambda}\otimes\tilde{H}_{\blambda}(t)\right)&\ket{\blambda'}_A\otimes \ket{\tilde{\varepsilon}_{n,\blambda'}(t)}_P\nonumber \\
    =&\tilde{\varepsilon}_{n,\blambda'}(t)\ket{\blambda'}_A\otimes \ket{\tilde{\varepsilon}_{n,\blambda'}(t)}_P\nonumber,\\
    H(t)\mathbb{W}\ket{\blambda'}_A\otimes \ket{\tilde{\varepsilon}_{n,\blambda'}(t)}_P=&\tilde{\varepsilon}_{n,\blambda'}(t)
    \mathbb{W}\ket{\blambda'}_A\otimes \ket{\tilde{\varepsilon}_{n,\blambda'}(t)}_P.
\end{align}

This indicates that the eigenstates of $H(t)$ are $\mathbb{W}\ket{\tilde{\varepsilon}_{n,\blambda}(t)}_P\otimes\ket{\blambda}_A$ and eigenvalues are $\tilde{\varepsilon}_{n,\blambda}(t)$, because the number of $\tilde{\varepsilon}_{n,\blambda'}(t)$ is the same as that of eigenvalues of $H(t)$.

\section{Symmetry Preserving in Lindblad Equation}
We consider the case where the transversal couplings are negligible $|g^z_{i,k}|\gg|g^x_{i,k}|\sim0$ $\forall i,k$. We transform \eqref{eq:master} and \eqref{eq:lind} using $\mathbb{W}$: 
\begin{align}
    \dot{\tilde{\rho}}(t)=&-i[\tilde H(t),\tilde \rho(t)]+\sum_{\omega}\sum_{i,j=1}^{2N} g^z_{i,|\omega|} g^z_{j,|\omega|}\gamma(\omega)\nonumber\\
    \times &\biggl[\tilde{L}_{j ,\omega}(t)\tilde\rho(t)\tilde{L}^\dagger_{i,\omega}(t)-\frac{1}{2}\biggl\{\tilde{L}^\dagger_{i,\omega}(t)\tilde{L}_{j,\omega}(t),\tilde{\rho}(t)\biggr\}\biggr], \label{eq:master_2}\\
    \tilde{L}_{i,\omega}(t)=&\sum_{\omega_{ba}(t)=\omega}\ket{\tilde E_a(t)}\bra{\tilde E_a(t)}\mathbb{W}^\dagger\sigma_i^z\mathbb{W}\ket{\tilde E_b(t)}\bra{\tilde E_b(t)}, \label{eq:lind_2}
\end{align}
where we defined $\tilde{\rho}(t)=\mathbb{W}^\dagger \rho(t)\mathbb{W}$ and the eigenvalues and eigenstates of $\tilde{H}(t)$ as $\ket{\tilde{E}_a(t)}$ and $\tilde{E}_a(t)$. If the initial state is $\ket{\tilde\psi(0)}=\bigotimes_{i=1}^N\ket{1}_{2i-1}\ket{+}_{2i}$, the dynamics is confined to a subspace because the eigenstates of $\tilde{H}(t)$ are decomposed:
\begin{align}
    \ket{\tilde{E}_a(t)}=\ket{\blambda}_A\otimes\ket{\tilde\varepsilon_{n,\blambda}(t)}_P,
\end{align}
where $a$ corresponds to $(\blambda,n)$. This allows us to transform \eqref{eq:master_2} into 
\begin{align}
    \dot{\tilde{\rho}}_{\vec{1}}(t)=&-i[\tilde H_{\vec{1}}(t),\tilde \rho_{\vec{1}}(t)]\nonumber\\
    &+\sum_{\omega}\sum_{i,j=1}^{N} (g_{2i,|\omega|}^z-g_{2i-1,|\omega|}^z)(g_{2j,|\omega|}^z-g_{2j-1,|\omega|}^z)\nonumber\\
    &\times\gamma(\omega) \biggl[\tilde{L}_{2j ,\omega,\vec{1}}(t)\tilde\rho_{\vec{1}}(t)\tilde{L}^\dagger_{2i,\omega,\vec{1}}(t)\nonumber \\
    &\hspace{10mm}-\frac{1}{2}\biggl\{\tilde{L}^\dagger_{2i,\omega,\vec{1}}(t)\tilde{L}_{2j,\omega,\vec{1}}(t),\tilde{\rho}_{\vec{1}}(t)\biggr\}\biggr], \label{eq:master_3}\\
    \tilde{L}_{i,\omega,\vec{1}}(t)=&\sum_{\omega_{ba}(t)=\omega}\ket{\tilde \varepsilon_{a,\vec{1}}(t)}\bra{\tilde \varepsilon_{a,\vec{1}}(t)}\sigma_i^z\ket{\tilde \varepsilon_{b,\vec{1}}(t)}\bra{\tilde \varepsilon_{b,\vec{1}}(t)}, \label{eq:lind_3} 
\end{align}
where $\blambda=\vec{1}$ is defined as $\blambda=(1,1,\cdots,1)$ and we defined
\begin{align}
    \tilde\rho(t)=\ket{\vec{1}}_A\bra{\vec{1}}\otimes \tilde{\rho}_{\vec{1}}(t).
\end{align}

\end{document}